\documentclass[trackchanges, twocolumn]{aastex701}

\newcommand\aastex{AAS\TeX}

\usepackage{subcaption}

\begin{document}

\title{Complex Energy-Dependent Behaviour of Quasi-Periodic Oscillation Observed in GRS 1915+105}

\author[orcid=0009-0003-1029-5201,sname='Sharma']{Vaibhav Sharma}
\affiliation{Department of Physics, Indian Institute of Technology Kanpur, Kanpur, Uttar Pradesh - 208016, India}
\email[show]{svbhv@iitk.ac.in,vaibhavsharmaiitk@gmail.com}

\author[gname=Ranjeev, sname='Misra']{Ranjeev Misra}
\affiliation{Inter-University Center for Astronomy and Astrophysics, Ganeshkhind, Pune, Maharashtra - 411007, India}
\email{rmisra@iucaa.in}

\author[gname=V., sname='Jithesh']{V. Jithesh}
\affiliation{Centre of Excellence in Astronomy and Astrophysics (CEAA), Department of Physics and Electronics, Christ University, Hosur Main Road, Bengaluru - 560029, India}
\email{jithesh.v@christuniversity.in}

\author[gname=Shivani,sname=Chaudhary]{Shivani Chaudhary}
\affiliation{Department of Physics, Banasthali Vidyapith, Rajasthan - 304022, India}
\email[]{shivani.chaudhary.research@gmail.com}

\author[gname=Khushi,sname=Jirawala]{Khushi Jirawala}
\affiliation{Indian Institute of Astrophysics, Bengalore, India}
\email[]{khushijirawala@gmail.com}

\author[gname=J S,sname=Yadav]{J S Yadav}
\affiliation{Space, Planetary and Astronomical Sciences and Engineering, IIT Kanpur, Kanpur Nagar, Uttar Pradesh - 208016, India}
\affiliation{Department of Astronomy and Astrophysics, Tata Institute of Fundamental Research, Mumbai, Maharashtra - 400005, India}
\email{jsyadav@iitk.ac.in}

\author{Pankaj Jain}
\affiliation{Space, Planetary and Astronomical Sciences and Engineering, IIT Kanpur, Kanpur Nagar, Uttar Pradesh - 208016, India}
\email{pkjain@iitk.ac.in}

\author[gname=N J,sname=Juris]{N J Juris}
\affiliation{St.Thomas College, Ranni, Pathanamthitta—689673, India}
\email[]{njjuris@gmail.com}

\begin{abstract}
We present complex energy-resolved properties of quasi-periodic oscillations (QPOs) in the black hole X-ray binary GRS 1915+105 using an observation from the \texttt{LAXPC} instrument onboard AstroSat. Power density spectra (PDSs) are constructed in multiple energy bands and modeled with multi-Lorentzian components to investigate the energy dependence of QPO properties. The QPO frequency shows a modest increase with energy. Dynamic PDS analysis does not reveal clear evidence for time-dependent evolution of the QPO frequency, suggesting that the observed frequency shift is not primarily driven by temporal variability. We perform simultaneous fitting of energy-resolved PDSs and find that a model in which the QPO feature is described by two Lorentzian components provides a better fit. The two components exhibit different evolution in fractional root mean square amplitude as a function of energy. We further examine the phase-lag properties by simultaneously modeling the PDS and the real and imaginary parts of the cross-spectrum and find distinct phase-lag behavior for the two components. Overall, these results indicate that the apparent energy-dependent evolution of the QPO feature may be a result of the presence of more than one variability component.
\end{abstract}



\section{Introduction} \label{Section: Introduction} 
Accreting black hole X-ray binaries (BHXBs) exhibit rich variability on a wide range of timescales, providing key insights into the physics of accretion and the geometry of the inner flow \citep[e.g.,][and references therein]{Levine_2006, Remillard&McClintock_2006, Done_2007, Belloni&Motta_2011, Ingram&Motta_2019}. Among these variability features, quasi-periodic oscillations (QPOs) (\citealt{Nowak_2000, Belloni_2002}; for reviews, see
\citealt{Remillard&McClintock_2006, Ingram&Motta_2019}), are of particular interest, as they are thought to originate in the innermost regions of the accretion flow and therefore carry information about processes near the black hole \citep[e.g.,][]{van_der_Klis_1989, Psaltis&van_der_Klis_1999}.

Low-frequency QPOs (LFQPOs) (\citealt{Casella_2005}; for reviews, refer to \citealt{Remillard&McClintock_2006, Ingram&Motta_2019}), typically observed in the range of $\sim 0.1-30$ Hz, are commonly detected in BHXBs during their different spectral states. These QPOs are often classified into types A, B, and C based on their timing and spectral properties \citep[e.g.,][]{Wijnands_1999, Sobczak2000, Motta_2015, Motta_2016, Ingram&Motta_2019}. Despite extensive observational and theoretical efforts, the physical origin of LFQPOs remains unclear. Proposed models include geometric effects such as Lense–Thirring precession of a misaligned inner accretion flow \citep[e.g.,][]{Stella&Vietri_1998, Stella_1999, Ingram&Done_2009}, as well as intrinsic oscillation mechanisms involving the accretion flow, including transition-layer oscillations \citep{Titarchuk_2000}, shock oscillations within the two-component advective flow framework \citep{Chakrabarti_2008}, and oscillatory variability of the Comptonizing region inferred from spectral–timing studies \citep{Karpouzas_2020, Bellavita_2022}.

The energy-dependent properties of LFQPOs, including the fractional rms ($frms$) amplitude and phase lag, provide important constraints on the geometry and radiative processes of the accretion flow \citep{vanderKlis_1985, Nowak_1997b, Nowak_1999, Rodriguez_2004, Sobolewska_2006, Karpouzas_2020}. The $frms$ amplitude of LFQPOs increases with energy, often flattening or turning over at higher energies, suggesting that the variability is predominantly associated with the Comptonized emission rather than the thermal disk \citep[e.g.,][]{Rodriguez_2004, Casella_2004, Sobolewska_2006, Karpouzas_2020, Sharma_2026}. Likewise, phase-lag measurements across different energy bands provide important diagnostics of the underlying radiative processes. Both hard and soft lags have been observed in LFQPOs, with their origin commonly attributed to Comptonization, phase-dependent spectral evolution, or oscillations of the Comptonizing region \citep[e.g.,][]{Nowak_1997b, Nowak_1999, Qu_2010, Eijnden_2016, Karpouzas_2020, Bellavita_2022}.

An additional observational feature that has attracted considerable attention is the energy dependence of the QPO frequency \citep[e.g.,][]{Qu_2010, Li_2013, Li_2013b, Eijnden_2016, Eijnden_2017, Yan_2018, Zhu_2024, Mendez_2024}. Several studies have shown that the variation of QPO frequency with photon energy is not uniform but instead depends on the QPO frequency itself. Using RXTE observations of GRS 1915+105, \citet{Qu_2010} demonstrated that QPOs below $\sim$3 Hz exhibited a decrease in centroid frequency with increasing energy, whereas those above $\sim$3 Hz showed the opposite trend, with frequency increasing toward higher energies. Interestingly, the sign of the QPO phase lag has also been reported to change at approximately the same QPO frequency \citep[e.g.,][]{Reig_2000, Zhang_2020}. A similar but more complex behavior has been reported in XTE J1550--564, where \citet{Li_2013} found that the centroid frequency remained nearly independent of energy at the lowest frequencies ($\lesssim$0.4--0.8 Hz), showed only a weak increase up to $\sim$3.3 Hz, and displayed a clear positive correlation with energy at higher frequencies, with the strongest dependence observed around $\sim$6--8.5 Hz followed by a gradual flattening. More recently, \citet{Zhu_2024} reported comparable trends in Swift J1727.8-1613 using Insight-HXMT observations, where the centroid frequency was largely energy-independent at low frequencies ($\lesssim$3 Hz), increased with energy at higher frequencies, and showed a reduced rate of increase above $\sim$8 Hz.

While several studies, e.g., \cite{Qu_2010, Li_2013, Yan_2018} and \cite{Zhu_2024}, reported an apparent energy dependence of QPO centroid frequency, recent work by \cite{Mendez_2024} challenged their interpretation. Using a simultaneous fitting approach to the PDS and the real and imaginary components of the cross spectrum, \citet{Mendez_2024} showed no intrinsic energy dependence of the QPO frequency in GRS 1915+105 using one of the same RXTE observations reported by \cite{Qu_2010}. Instead, they discussed that the observed energy-dependent shifts could be attributed to the presence of two closely spaced variability components, named QPO and QPO shoulder, with different energy-dependent $frms$ amplitude and time-lag evolutions. A similar interpretation has recently been proposed for the black hole X-ray binary Swift J1727.8$-$1613, where the complex timing properties were interpreted as arising from the coexistence of Type-B and Type-C QPOs \citep{Jin_2026}.

The black hole X-ray binary GRS 1915+105 is one of the most variable and extensively studied accreting systems, exhibiting a rich diversity of timing behaviour \citep[e.g.,][and many more]{Morgan_1997, Muno_1999, Belloni_2000, Reig_2000, Rodriguez_2002, Ingram_2015, Yadav_2016, Rawat_2022}. Using observations from AstroSat, \cite{Yadav_2016} explored the high-energy variability of this source and presented, in their Figure 7, an intriguing indication that the QPO frequency evolved with energy, shifting from $\sim6.55$ Hz in the 3.0–8.0 keV band to $\sim7.48$ Hz in the 20.0–80.0 keV band during Orbit 02360 of observation \texttt{T01$\_$030T01$\_$9000000358}. Although this apparent energy-dependent shift in QPO frequency was not explicitly discussed or examined in detail in their work, it points to a potentially important aspect of the underlying variability mechanism. We refer to this feature as a QPO feature, as it may comprise two distinct variability components with different $frms$ amplitude and phase-lag behaviours. Although the observation \texttt{T01$\_$030T01$\_$9000000358} has also been analyzed in several other studies \citep{Banerjee_2021, Majumder_2022, Dhaka_2023, Belloni_2024}, none of them addressed this intriguing feature of an apparent energy dependence in the QPO frequency.

Motivated by this suggestive feature, we focus specifically on this orbit and undertake a dedicated investigation to probe the nature of the observed variability. In this work, we present detailed energy-dependent properties of the QPO feature in GRS 1915+105 using data from Orbit 02360 obtained with the LAXPC onboard AstroSat. We explore the energy dependence of QPO feature properties by constructing the PDSs across multiple energy bands and performing both independent and simultaneous modelling. Furthermore, we investigate the phase-lag behaviour using a framework that involves the joint modelling of the power spectrum and the real and imaginary parts of the cross-spectrum, following the methodology outlined in \cite{Mendez_2024}.

The structure of this paper is as follows. In Section \ref{Section: Observation and Data Reduction}, we describe the observations and data reduction procedures. Section \ref{Section: Analysis and Results} presents the analysis and results, where we first discuss the construction of power density spectra and the adopted data segmentation scheme (Section \ref{Subsec: PDS and Data Segmentation}), followed by a detailed investigation of the energy dependence of the QPO feature (Section \ref{Subsec: Energy-Dependence of QPO Frequency}). We then explore the possible co-existence of two distinct variability components \citep[QPO and QPO shoulder;][Section \ref{Subsec: Co-existence of Two Variability Components}]{Belloni_1997, Jonker_2000, Mendez_2024} and examine the behaviour of the $frms$ amplitude variability (Section \ref{Subsec: Fraction Root Mean Square Behaviour}). In Section \ref{Subsec: Phase Lag Calculation and Evolution}, we study the phase-lag evolution using the \texttt{constant phase lag model}, a methodology discussed in \cite{Mendez_2024}. Finally, we discuss and conclude the implications of our findings in Section \ref{Section: Discussion}.

\section{Observation and Data Reduction} \label{Section: Observation and Data Reduction}
AstroSat \citep[][]{Singh_2014}, the first Indian multiwavelength satellite, observed GRS 1915+105 multiple times since its launch. In this work, we utilize data from the Large Area Proportional Counter \citep[LAXPC;][]{Roy_2016, Yadav_2016, Yadav_2016a}, one of the onboard instruments, corresponding to a single orbit (02360) from the observation \texttt{T01$\_$030T01$\_$9000000358} carried out in March 2016. AstroSat/LAXPC is well known for its excellent timing capabilities, offering a time resolution of 10~$\mu$s, broad energy coverage in the 3.0–80.0 keV range, and a large effective area of $\sim$6000 cm$^{2}$ (combining all three detector units) at 15 keV.

After downloading the Level-1 data from the \href{https://astrobrowse.issdc.gov.in/astro_archive/archive/Home.jsp}{AstroSat
 data archive\footnote{\url{https://astrobrowse.issdc.gov.in/astro_archive/archive/Home.jsp}}}, we follow the standard procedures provided in the \href{https://www.tifr.res.in/~astrosat_laxpc/LaxpcSoft.html}{LAXPC
 software\footnote{\url{https://www.tifr.res.in/~astrosat_laxpc/LaxpcSoft.html}}} for data reduction. The raw Level-1 data are first processed into Level-2 event files, which contain the arrival time and energy information of individual photons, using the \texttt{laxpc$\_$make$\_$event} task.  

\begin{figure}
	\includegraphics[width=\columnwidth, height=5.2cm]{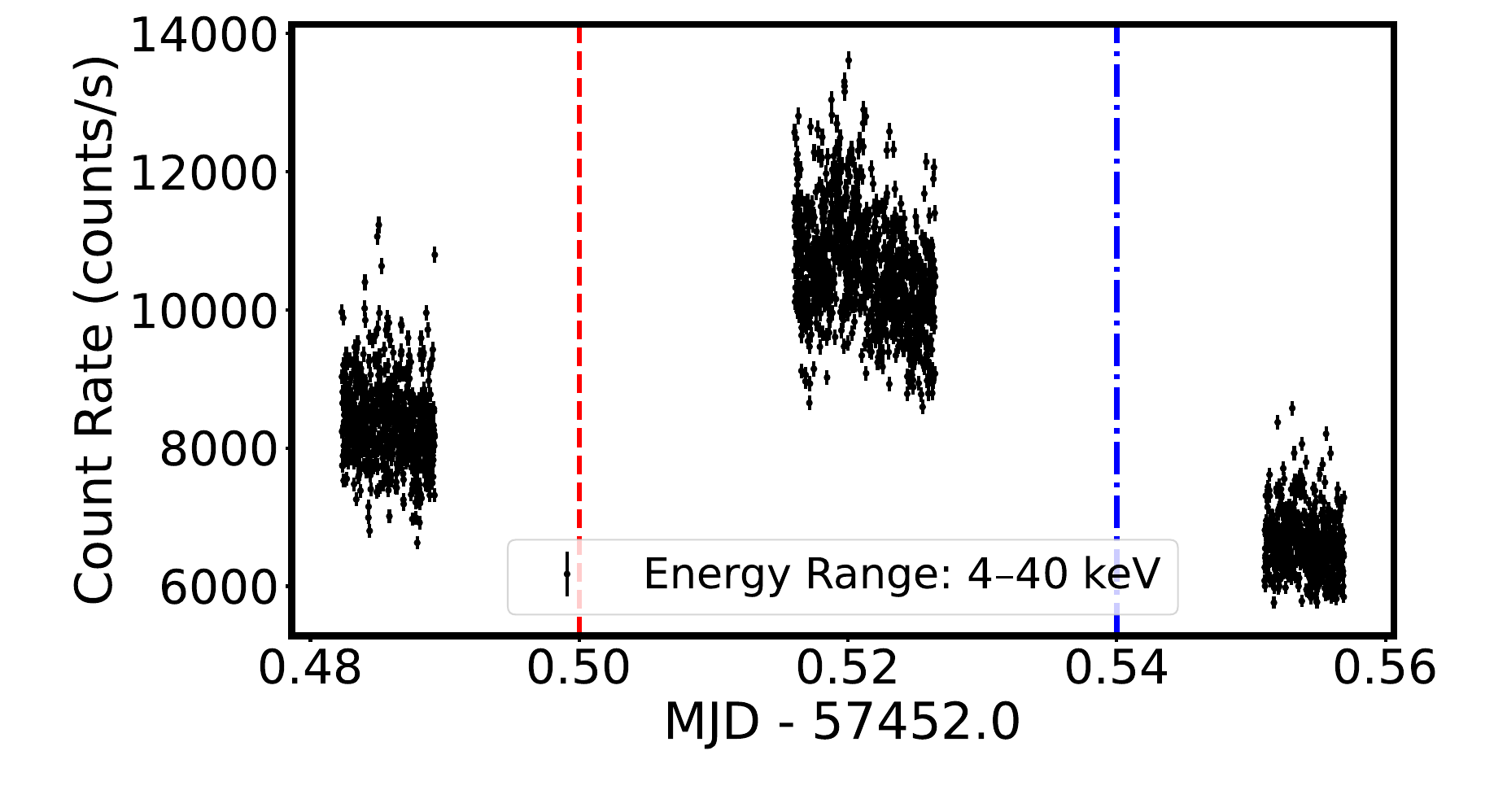}
    \caption{Orbit 02360 light curve of GRS 1915+105: The light curve is divided into three segments: Segment 1 (left of the red dashed line), Segment 2 (between the red dashed and blue dot-dashed lines), and Segment 3 (right of the blue dot-dashed line).}
    \label{Figure: Lightcurve}
\end{figure}

The PDS are generated using the \texttt{laxpc\_find\_freqlag} task. For the analysis of the energy-resolved PDSs presented in Sections \ref{Subsec: Energy-Dependence of QPO Frequency} and \ref{Subsec: Co-existence of Two Variability Components}, the Fourier frequency range is specified as 0.01-20 Hz. The task accordingly adopts a time resolution of 25~ms, corresponding to a Nyquist frequency of 20 Hz, and uses Fourier segments of 4096 time bins (102.4 s), yielding a minimum sampled frequency of $9.77 \times 10^{-3}$ Hz.

For the simultaneous modelling of the PDS and the cross-spectrum presented in Section \ref{Subsec: Phase Lag Calculation and Evolution}, the Fourier products are generated over a broader frequency range of 0.01-100 Hz. In this case, \texttt{laxpc\_find\_freqlag} automatically adopts a time resolution of 5 ms, corresponding to a Nyquist frequency of 100 Hz, and uses Fourier segments of 16384 time bins (81.92 s), yielding a minimum sampled frequency of $1.22\times10^{-2}$ Hz. The resulting PDS are subsequently rebinned in frequency using the \texttt{laxpc\_rebin\_power} routine, requiring a minimum signal-to-noise ratio of 3 while allowing the logarithmic frequency-bin width factor to vary between 1.01 and 1.05.

The PDS are expressed in background-corrected $frms$ normalization following the prescription of \cite{Belloni_1990}. The \texttt{laxpc$\_$find$\_$freqlag} task first subtracts the dead-time-corrected Poisson noise before computing the $frms$ amplitude and subsequently applies the background correction factor, $C/(C-B)$, where $C$ and $B$ denote the total and background count rates, respectively. Thus, the adopted normalization is equivalent to the standard background-corrected $frms$ normalization described by \cite{Belloni_1990}.

\section{Analysis and Results} \label{Section: Analysis and Results}
\subsection{Power Density Spectrum and Data Segmentation} \label{Subsec: PDS and Data Segmentation}
\begin{figure*}
    \centering
    \includegraphics[angle=0, width=0.48\textwidth]{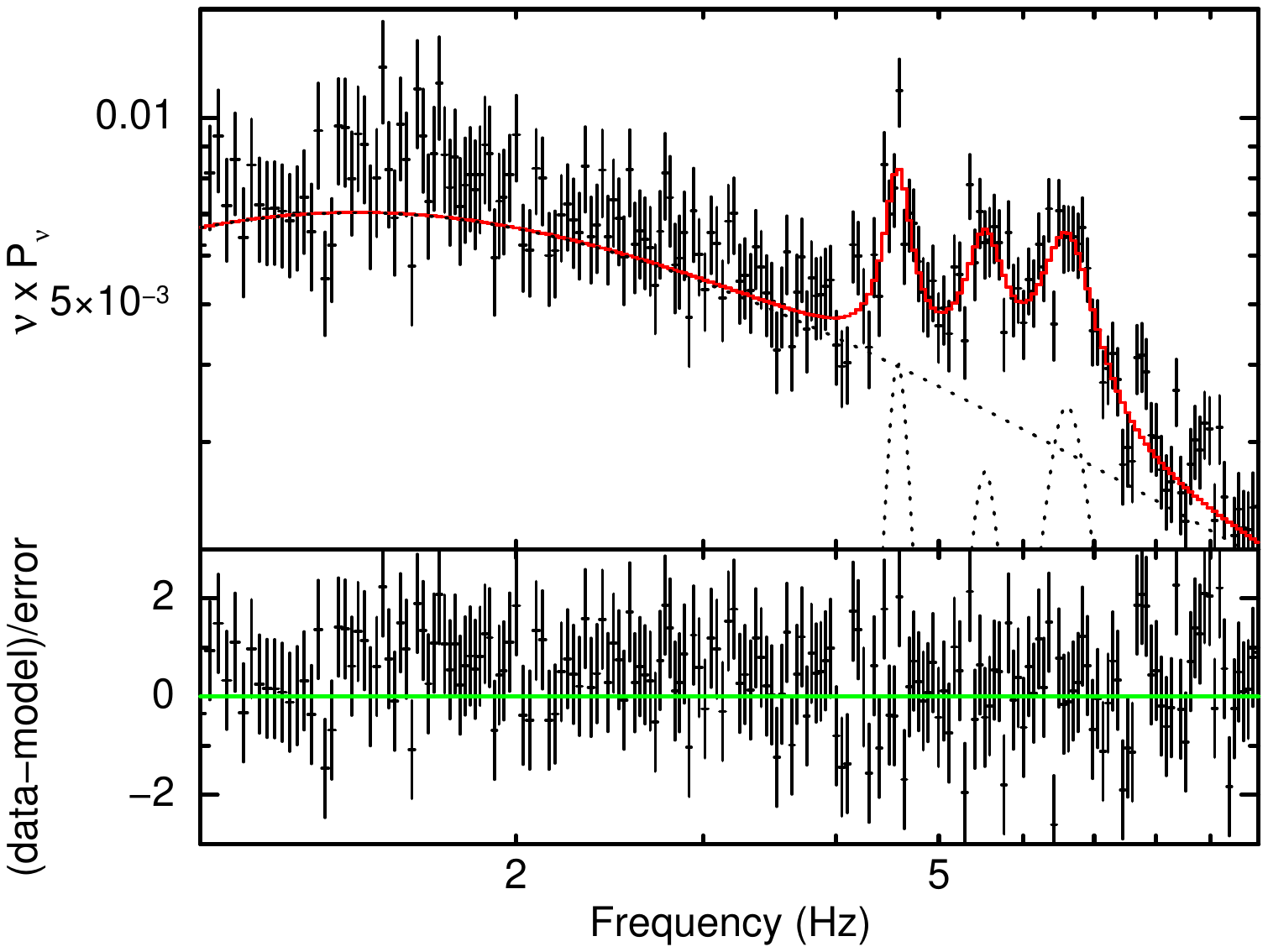}
    \includegraphics[angle=0, width=0.48\textwidth]{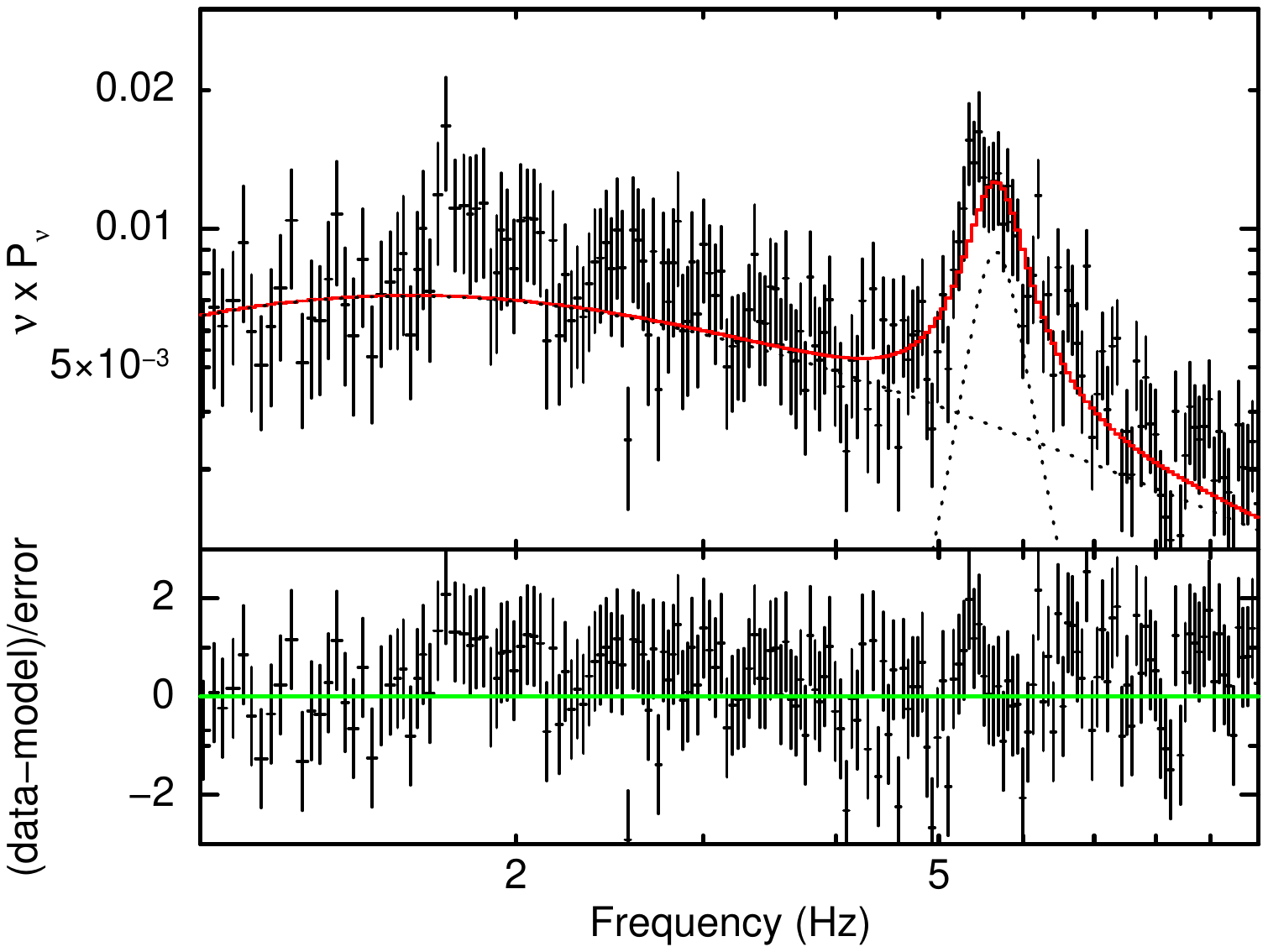}

    \includegraphics[angle=0, width=0.48\textwidth]{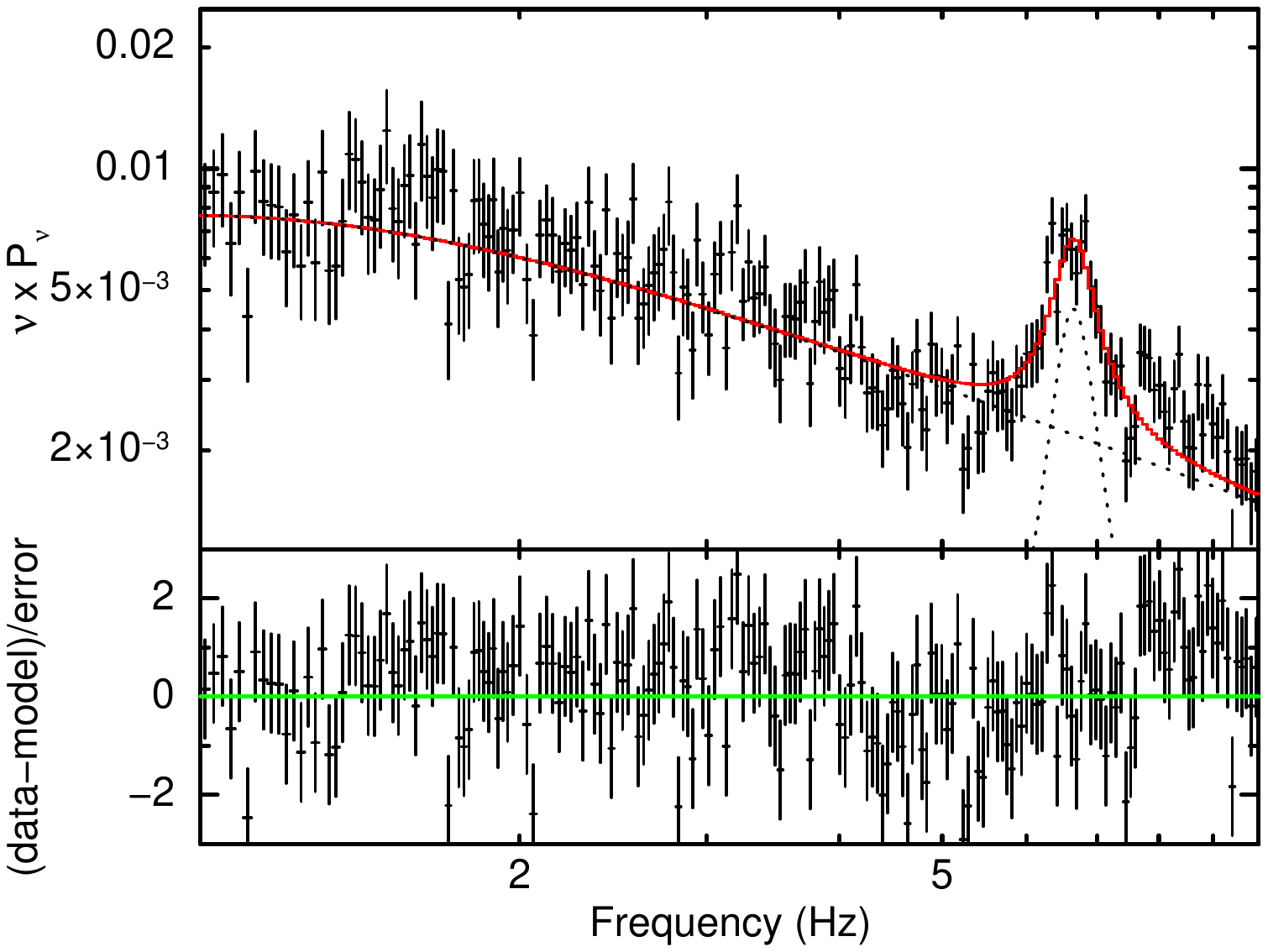}
    \includegraphics[width=0.48\textwidth]{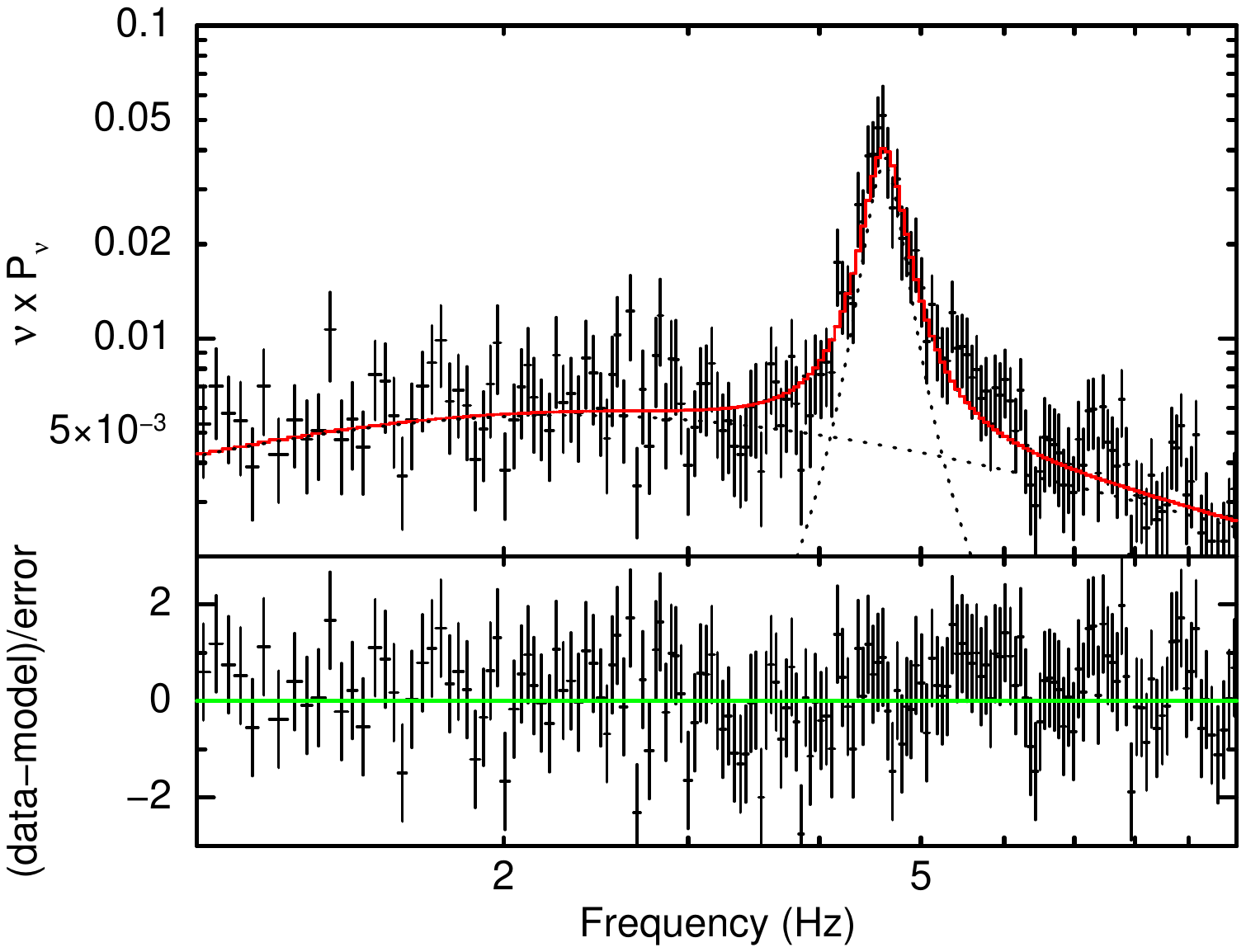}

    \caption{Power density spectra (PDS) of GRS 1915+105 in the 4.0-40.0 keV energy range. The figure shows the PDSs of the full orbit (top left) and Segments 1 (top right), 2 (bottom left), and 3 (bottom right). Bottom panels of all PDSs represent the residuals}.
    
    \label{Figure: PDS_2360_All}
\end{figure*}

We extract the background-subtracted light curve of orbit 02360 in the 4.0–40.0 keV energy range (Figure \ref{Figure: Lightcurve}) and generate the corresponding PDS (top left panel of Figure \ref{Figure: PDS_2360_All}) for the full orbit. The PDS exhibits three distinct peaks, as shown in the top left panel of Figure \ref{Figure: PDS_2360_All}. To model the PDS, we adopt the standard \texttt{multi-Lorentzian} approach \citep[e.g.,][]{Nowak_2000, Belloni_2002}. A combination of four Lorentzian components is required to obtain a statistically acceptable fit, yielding $\chi^2 = 398$ for 339 degrees of freedom (DOF). One of the Lorentzian components is fixed at zero centroid frequency to account for the zero-centred broad-band noise (BBN), while the centroid frequencies of the remaining three components are allowed to vary freely. The resulting centroid frequencies are $4.58\pm0.02$ Hz, $5.51\pm0.04$ Hz, and $6.57\pm0.06$ Hz. All uncertainties in the text and error bars in the figures are reported at the one sigma confidence level in this work.

Further, we subdivide the light curve into three segments (see Figure \ref{Figure: Lightcurve} for segmentation details) to investigate whether the observed frequencies exhibit any time dependence. For each segment, we generate the corresponding PDS and fit it using a multi-Lorentzian model. We find that each segmental PDS can be fitted by two Lorentzian components, yielding statistically acceptable fits. Notably, each PDS exhibits a single peaked feature, with centroid frequencies at $5.64\pm0.04$ Hz (Figure \ref{Figure: PDS_2360_All}; top right panel), $6.64\pm0.03$ Hz (Figure \ref{Figure: PDS_2360_All}; bottom left panel), and $4.61\pm0.02$ Hz (Figure \ref{Figure: PDS_2360_All}; bottom right panel) for Segments 1, 2, and 3, respectively. The presence of only one peak in each segmental PDS suggests that the three-peaked structure observed in the PDS of the full orbit is likely a result of averaging of PDS with a single Lorentzian whose frequency changes.

\subsection{Energy-Dependence of QPO Feature} \label{Subsec: Energy-Dependence of QPO Frequency}
\begin{figure}
	\includegraphics[width=\columnwidth]{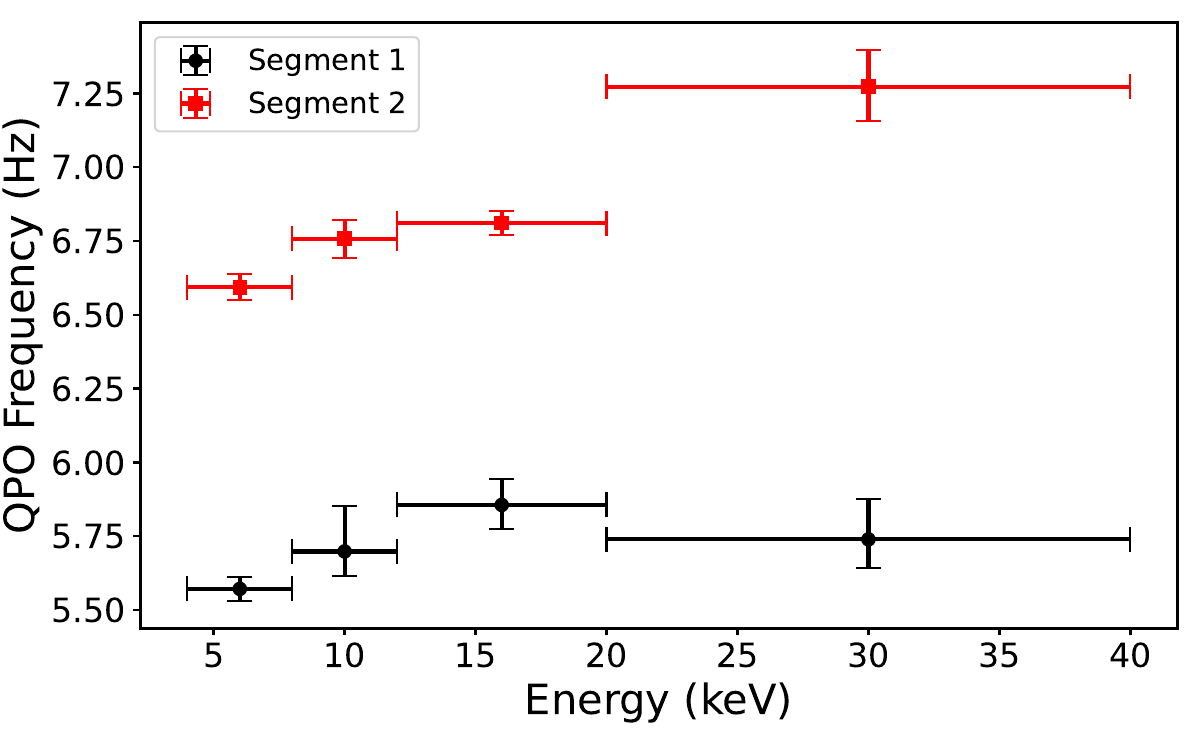}
    \caption{Energy Dependence of QPO Frequency in GRS 1915+105: The red and black data points correspond to Segments 1 and 2, respectively.}
    \label{Figure: QPO_Variation_with_Energy}
\end{figure}

\begin{figure*}
    \centering
    \includegraphics[width=0.48\textwidth, height=5.5cm]{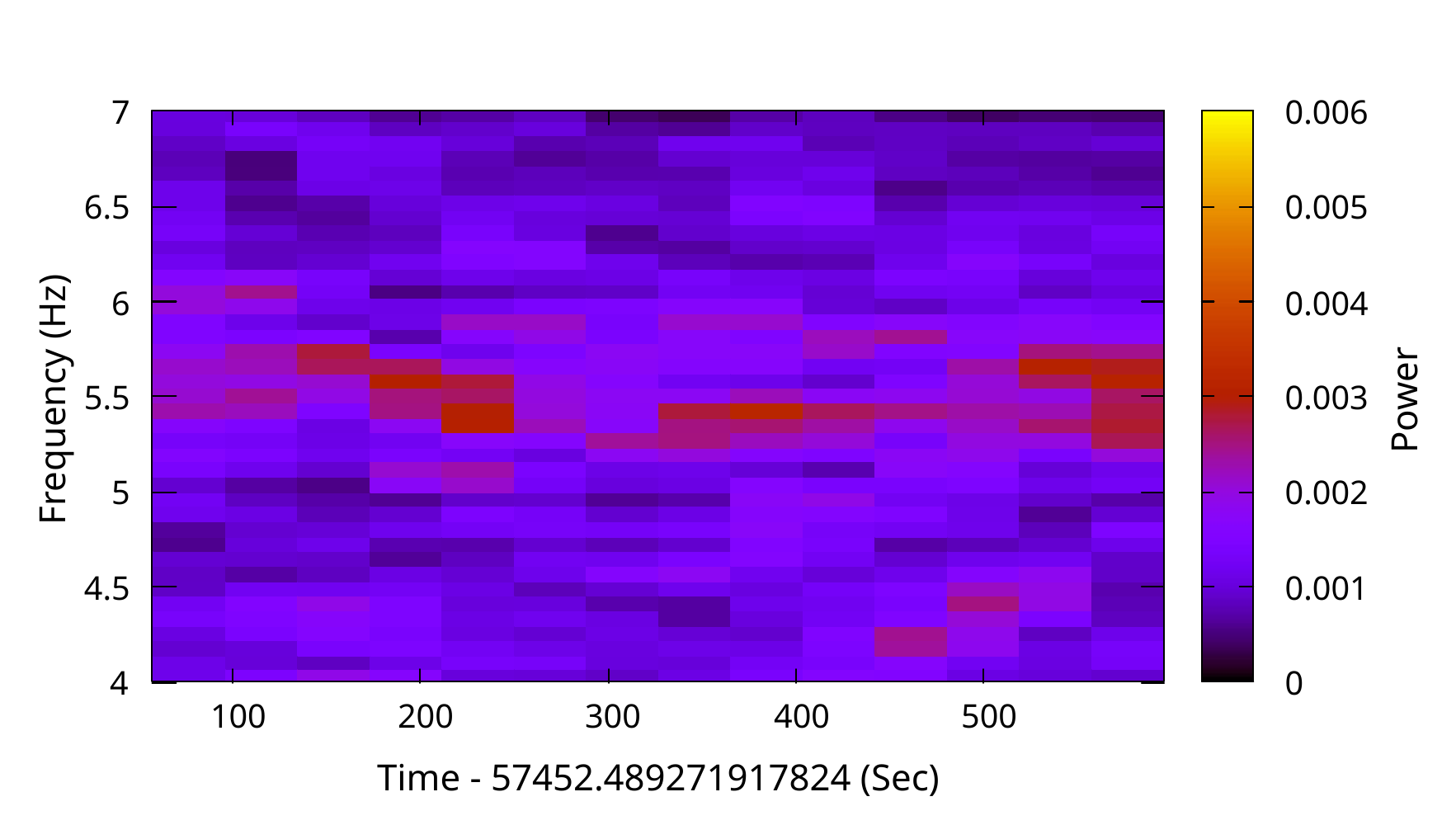}
    \includegraphics[width=0.48\textwidth, height=5.5cm]{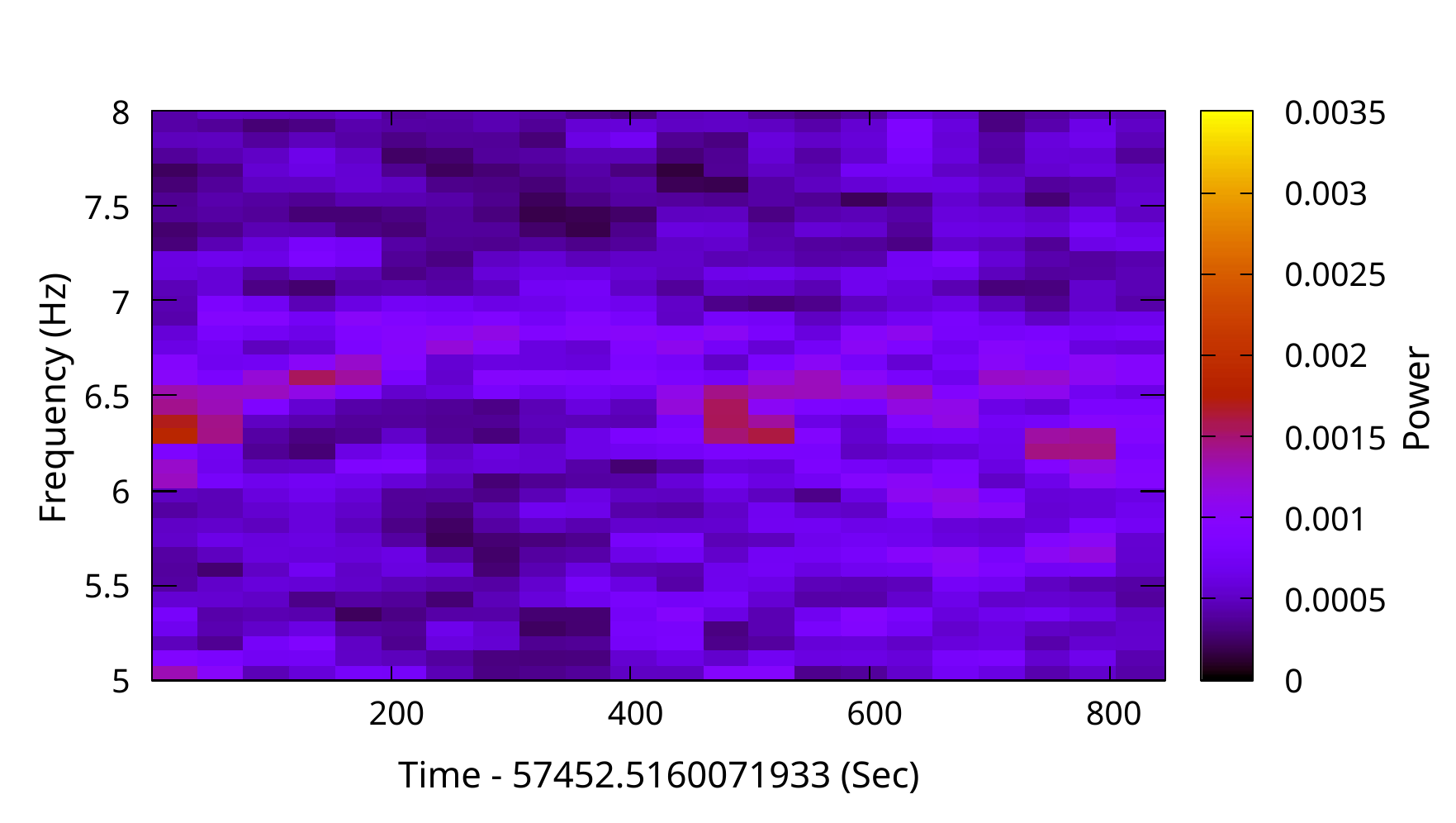}
    
    \caption{Dynamic power density spectra of Segments 1 (left) and 2 (right) of GRS 1915+105}.
    
    \label{Figure: DPDS_Segments_1_2}
\end{figure*}

Next, we investigate the energy dependence of the QPO feature in each segment. We generate PDSs in four energy bands: 4.0–8.0 keV, 8.0–12.0 keV, 12.0–20.0 keV, and 20.0–40.0 keV and fit them simulataneously using a \texttt{multi-Lorentzian} model (specifically two Lorentzians: one representing the zero-centred BBN and the other representing the QPO feature). The centroid frequency of the BBN component is fixed at zero, while all other parameters are allowed to vary freely and independently across the four energy-resolved PDSs. This model provides an acceptable fit, with a $\chi^2$ of 471.1 for 509 degrees of freedom in Segment 1 and a $\chi^2$ of 714.7 for 660 degrees of freedom in Segment 2. We find that the QPO feature exhibits a systematic shift from $5.57\pm0.04$ Hz (4.0-8.0 keV) to $5.86\pm0.08$ Hz (12.0-20.0 keV) in Segment 1 (Figure \ref{Figure: QPO_Variation_with_Energy}; black data points) and from $6.59\pm0.04$ Hz (4.0-8.0 keV) to $7.27\pm0.12$ Hz (20.0-40.0 keV) in Segment 2 (Figure \ref{Figure: QPO_Variation_with_Energy}; red data points), whereas it remains consistent across energy bands in Segment 3 (centroid frequency $\sim4.5$ Hz). To verify that the observed energy-dependent frequency shift is not driven by temporal evolution, we construct dynamic PDSs for Segments 1 and 2. The dynamic PDSs are generated using the \texttt{laxpc$\_$dynpower} task within the \texttt{LAXPC} software package in the 4.0-40.0 keV energy range. We adopt a segment length of $\sim$38 seconds and a frequency resolution of 0.08 Hz for both segments. The resulting dynamic PDSs (Figure~\ref{Figure: DPDS_Segments_1_2}) do not show any significant time-dependent variation in the QPO centroid frequency in either segment, suggesting that the observed frequency shift with energy is not time-dependent.

\subsection{Co-existence of Two Variability Components} \label{Subsec: Co-existence of Two Variability Components}
\begin{figure}
    \includegraphics[angle=0, width=1.1\columnwidth, height=6cm]{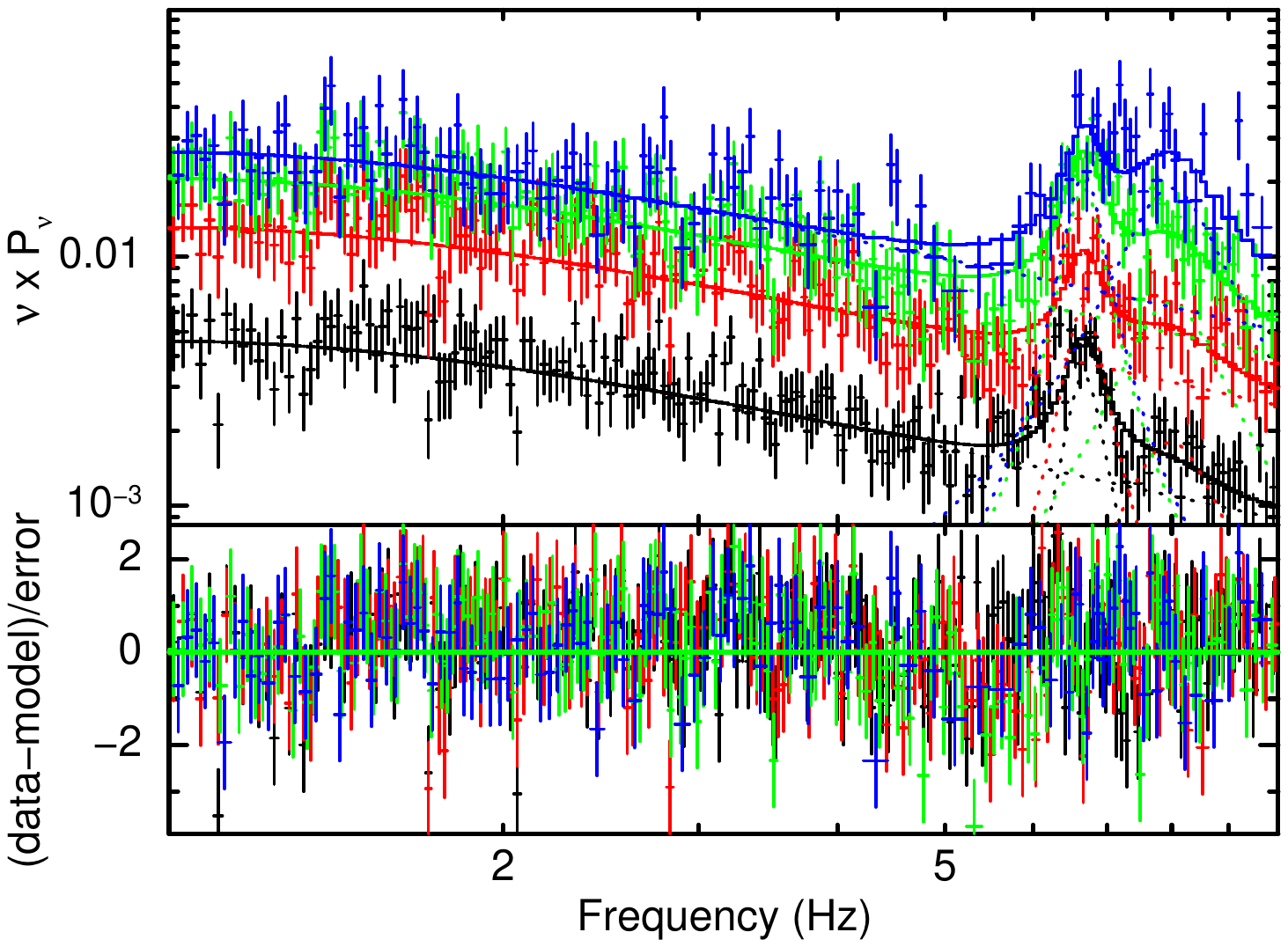}
    \caption{Simultaneous fits to the energy-resolved PDSs of Segment 2 of GRS 1915+105. The black, red, green, and blue points correspond to the 4.0–8.0 keV, 8.0–12.0 keV, 12.0–20.0 keV, and 20.0–40.0 keV bands, respectively. The PDSs are fitted with three Lorentzian components, with centroid frequencies and widths tied across all energy-resolved PDSs. Bottom panel represents the residual}.
    \label{Figure: PDS with 2 Lore Segment 2}
\end{figure}

To examine whether the QPO feature can be better described by two \texttt{Lorentzian} components \citep[QPO and QPO shoulder;][]{Belloni_1997, Jonker_2000, Mendez_2024}, we introduce an additional \texttt{Lorentzian} and perform simultaneous fits to all PDSs within each segment. In this case, the centroid frequency and width of each \texttt{Lorentzian} are tied across the PDSs in different energy bands, while their normalizations are allowed to vary independently. First, the centroid frequency and width of one \texttt{Lorentzian} are fixed to the QPO feature parameters obtained in the 4.0–8.0 keV band, and those of the second \texttt{Lorentzian} are fixed to the corresponding values from the 20.0–40.0 keV band. After the first fit, these parameters are allowed to vary freely but remain tied across the energy-resolved PDSs. This approach yields a smaller $\chi^2$ in both segments. For Segment 1, this approach provides a moderately improved fit, reducing $\chi^2$ from 471.1 to 445.5 for the same number of degrees of freedom. Similarly, for Segment 2, it provides a better fit, reducing $\chi^2$ from 714.7 to 671 for 660 degrees of freedom in both cases (Figure \ref{Figure: PDS with 2 Lore Segment 2}). We find that the observed feature is better described by two \texttt{Lorentzian} components, with centroid frequencies $5.52\pm0.04$ \citep[QPO with quality factor ($Q \sim13.8$);][]{van_der_Klis_1994} and $6.20\pm0.15$ (QPO shoulder with $Q\sim5.34$) in Segment 1, and $6.66\pm0.03$ (QPO with $Q\sim11.68$ ) and $7.94\pm0.12$ (QPO shoulder with $Q\sim4.64$) in Segment 2. The quality factor ($Q$) is defined as the ratio of the centroid frequency to the width of the \texttt{Lorentzian} profile.

\subsection{Fraction Root Mean Square Amplitude} \label{Subsec: Fraction Root Mean Square Behaviour}
\begin{figure*}
    \centering
    \includegraphics[width=0.48\textwidth]{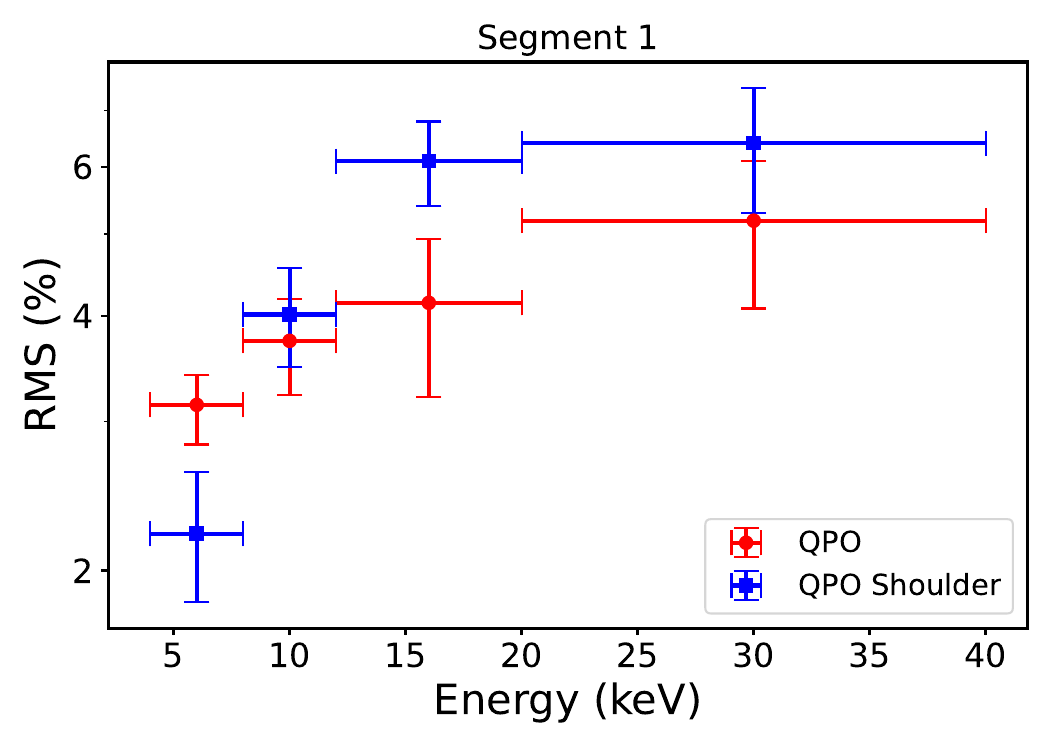}
    \includegraphics[width=0.48\textwidth]{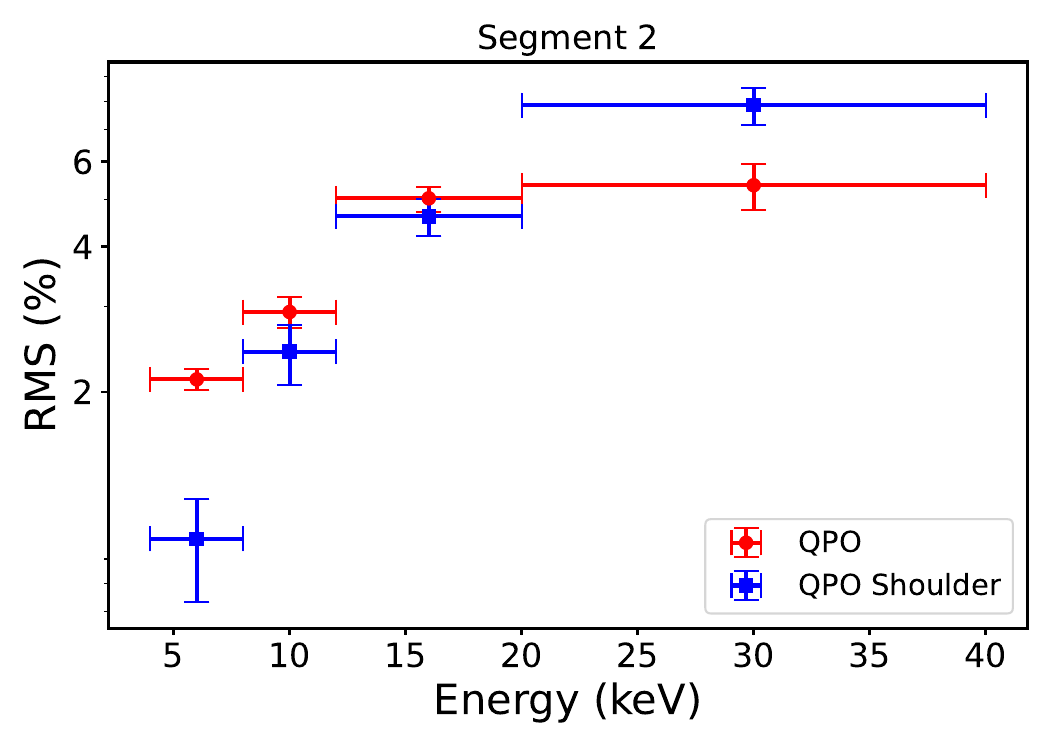}
    \caption{Energy dependence of $frms$ amplitude for Segments 1 (left) and 2 (right) of GRS 1915+105}.
    \label{Figure: RMS_Segments_1_2}
\end{figure*}

To investigate this further, we compute the $frms$ amplitude associated with the two \texttt{Lorentzian} components that describe the QPO feature in the PDSs of both segments. The $frms$ amplitude is estimated by taking the square root of the normalization of the respective \texttt{Lorentzian}, as the PDSs are normalized in units of $frms^2$. We then plot $frms$ amplitude as a function of energy in Figure \ref{Figure: RMS_Segments_1_2} for Segments 1 (left panel) and 2 (right panel), respectively.

For Segment 1 (Figure \ref{Figure: RMS_Segments_1_2}, left panel), the $frms$ amplitude of the QPO increases from $3.14 \pm 0.18\%$ in the 4-8 keV band to $5.18 \pm 0.99\%$ in the 20-40 keV band. The QPO shoulder exhibits a steeper increase, rising from $2.21 \pm 0.39\%$ to $6.41 \pm 1.07\%$ over the same energy range. Although the shoulder attains a higher $frms$ amplitude at high energies, the difference between the two components remains statistically insignificant in individual energy bands, reaching only $\sim0.8\sigma$ in the 20-40 keV band. Thus, Segment 1 shows only marginal evidence that the two variability components possess different $frms$ amplitude spectra.

In contrast, Segment 2 (Figure \ref{Figure: RMS_Segments_1_2}, right panel) exhibits a clearer distinction between the two components. The $frms$ amplitude of the QPO increases from $2.12 \pm 0.11\%$ in 4-8 keV to $5.36 \pm 0.58\%$ in 20-40 eV, whereas the QPO shoulder rises more rapidly from $0.99 \pm 0.23\%$ in 4-8 keV to $7.86 \pm 0.69\%$ in 20-40 keV. In the highest-energy band (20-40 keV), the $frms$ amplitude of the QPO shoulder exceeds that of the QPO by $2.50 \pm 0.90\%$, corresponding to a $\sim2.8\sigma$ difference. These results quantitatively demonstrate that the QPO shoulder exhibits a significantly steeper energy dependence of the $frms$ amplitude than the QPO.

\subsection{Phase Lag} \label{Subsec: Phase Lag Calculation and Evolution}
We further investigate the phase-lag behaviour of the QPO and QPO shoulder in both segments using the \texttt{constant phase lag model} approach presented by \cite{Mendez_2024}. In this framework, the variability is described as a superposition of Lorentzian components that are internally coherent while being mutually incoherent with one another. A key assumption of this model is that each variability component is associated with a linear response between energy bands, characterized by a complex transfer function whose phase is independent of Fourier frequency. Under this assumption, the cross-spectrum can be decomposed into contributions from individual components, where the real and imaginary parts are proportional to the same Lorentzian profile that describes the PDS, scaled by $\cos(\phi)$ and $\sin(\phi)$, respectively, with $\phi$ representing the phase lag of that component. This formulation provides a direct link between the power spectrum and the cross-spectrum (CS), enabling a simultaneous and self-consistent fit to the PDS and the real and imaginary parts of the CS.

\begin{figure*}
	\includegraphics[width=\textwidth]{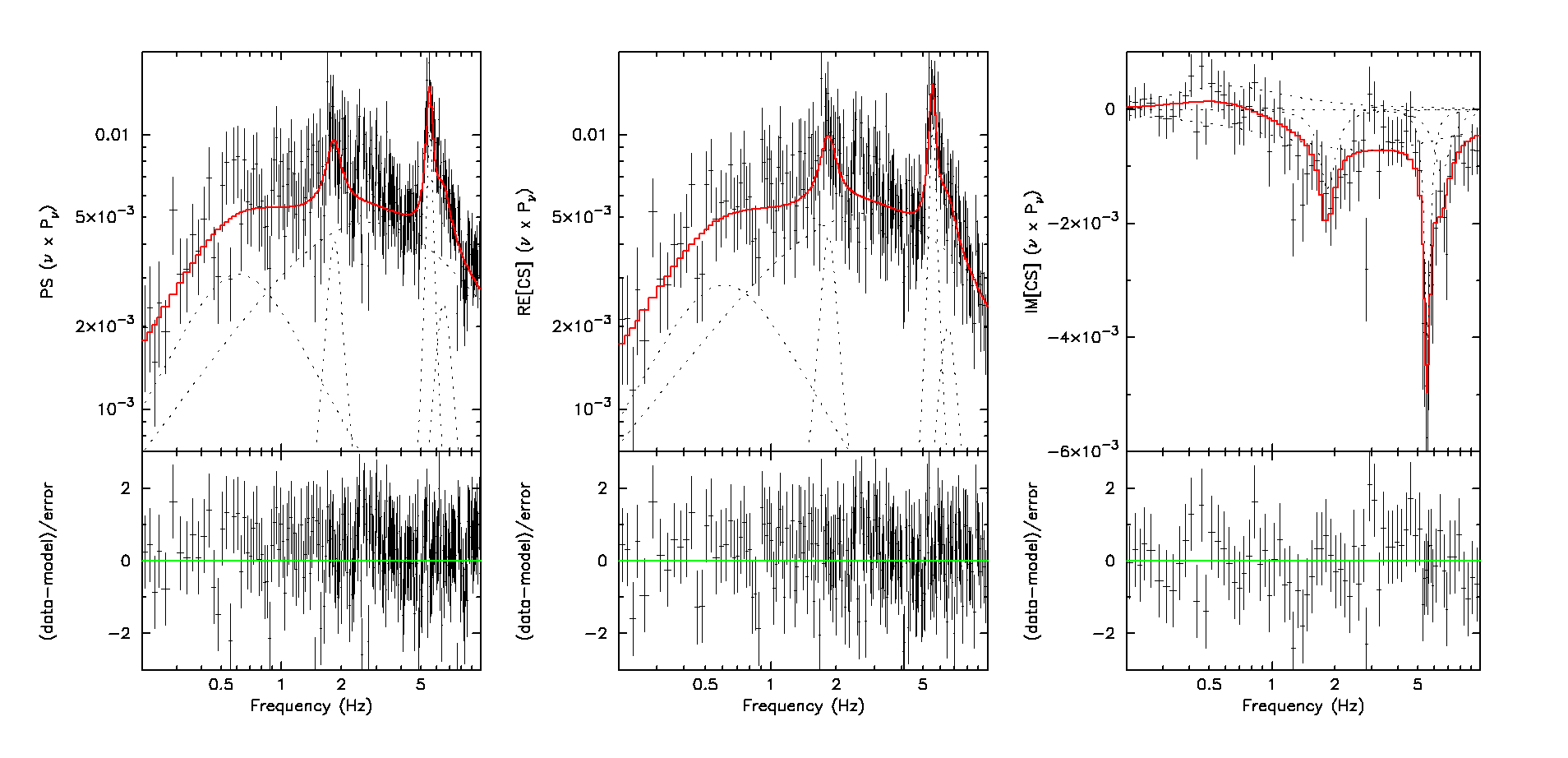}
    \caption{The top-left panel shows the fit (red solid line) to the PDS in the 4.0–40.0 keV energy range in Segment 1 of GRS 1915+105, modelled with five Lorentzian components (black dotted lines) and a power-law. The middle-top panel shows the fit to the real part of the cross-spectrum (8.0–12.0 keV relative to 4.0–8.0 keV), while the right-top panel shows the fit to the imaginary part of the cross-spectrum for the same energy bands. Bottom panels: The left, middle, and right panels show the corresponding $\chi$ residuals for the fits to the PDS and the real and imaginary parts of the cross-spectrum, respectively.}
    \label{Figure: Constant Phase Lag 1 Segment 1}
\end{figure*}

\begin{figure*}
	\includegraphics[width=\textwidth]{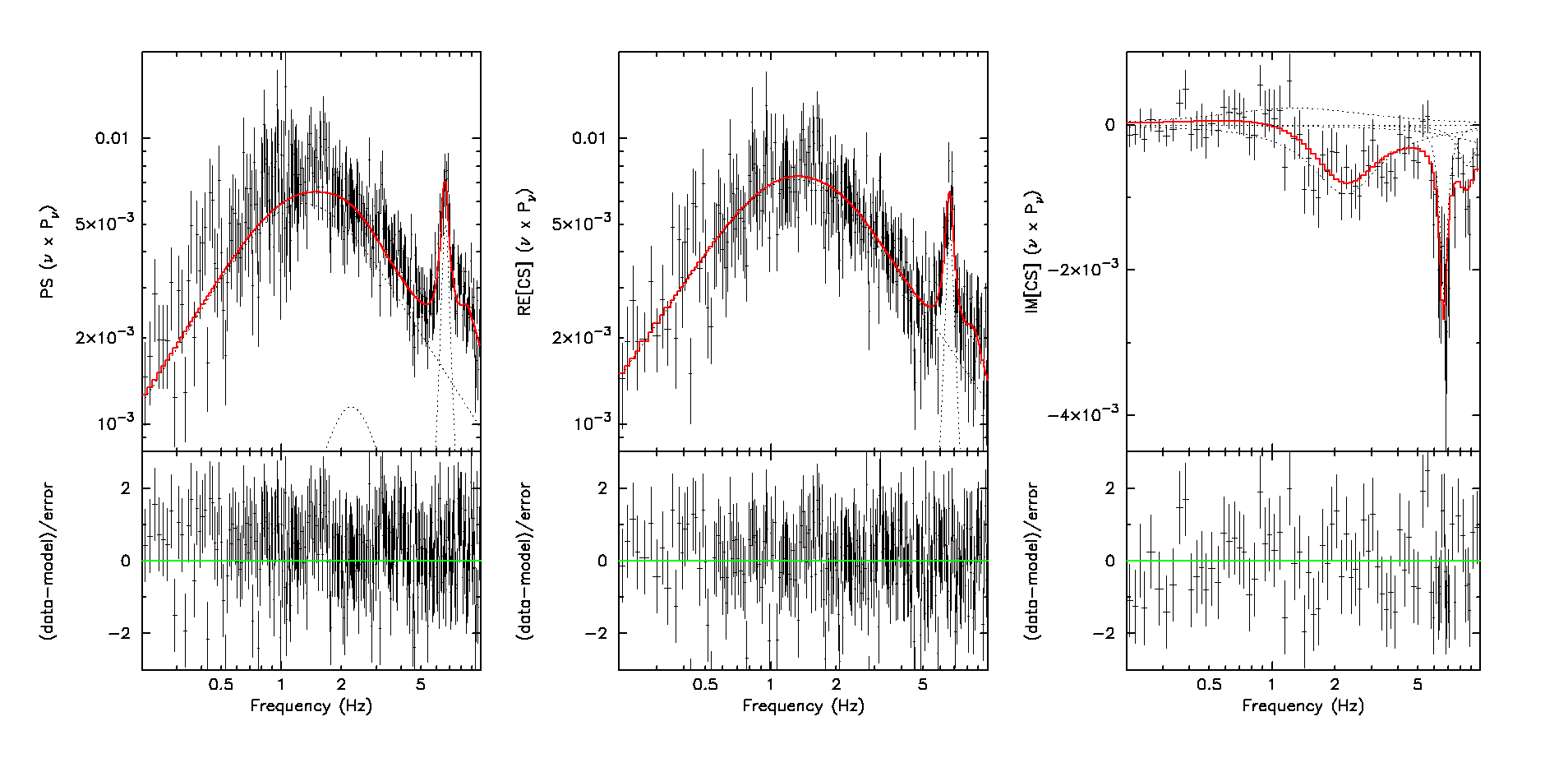}
    \caption{Representative simultaneous fits of PDS, real and imaginary part of the CS for Segment 2 of GRS 1915+105. Left, Middle, and Right: Same as Figure \ref{Figure: Constant Phase Lag 1 Segment 1}.}
    \label{Figure: Constant Phase Lag 1 Segment 2}
\end{figure*}

\begin{figure*}
	\includegraphics[width=0.95\textwidth]{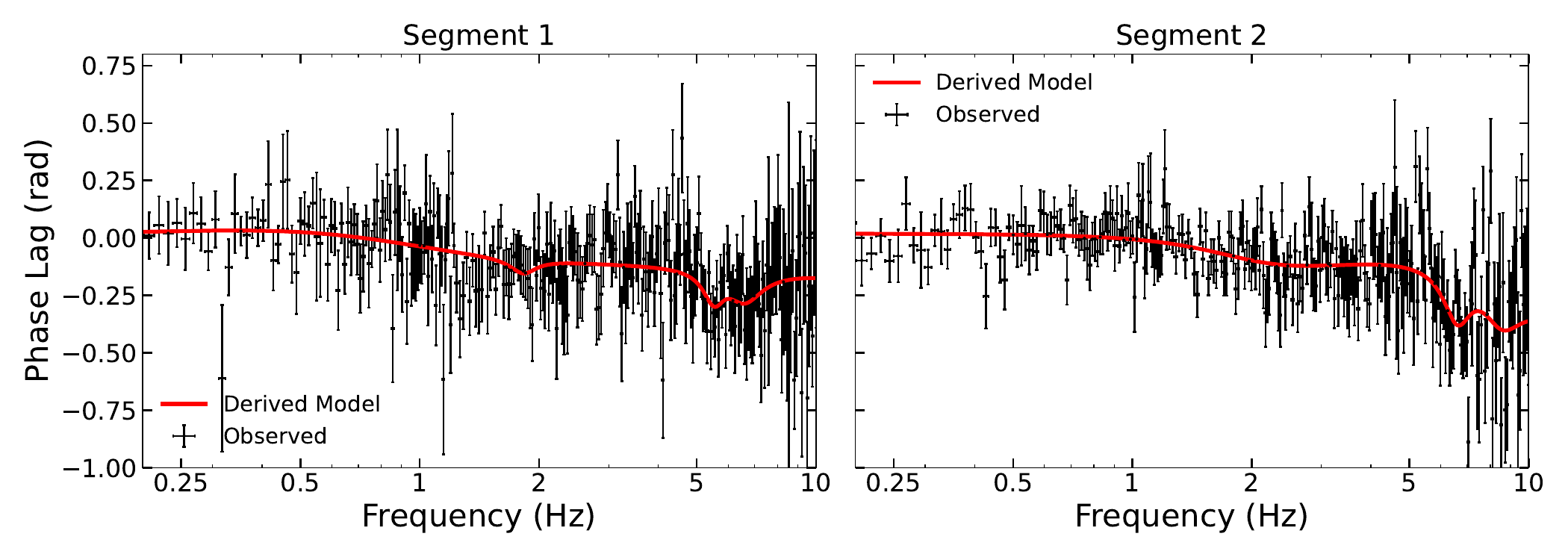}
    \caption{Observed and derived phase-lag spectra for Segments 1 (left panel) and 2 (right panel) of GRS 1915+105. The phase lags are calculated from the cross-spectrum between the 8.0-12.0 keV energy band and the 4.0-8.0 keV reference band. The black points with error bars represent the observed phase lags, calculated from the real and imaginary parts of the cross-spectrum. The solid red curves show the phase-lag spectra derived from the best-fitting models to the real and imaginary parts of the cross-spectrum, without directly fitting the phase lags.}
    \label{Figure: Phase Lag Spectrum}
\end{figure*}
Following this methodology, we first model the PDS in the full energy band (4.0–40.0 keV). The PDS is fitted by progressively adding \texttt{Lorentzian} components until no further improvement in $\chi^2$ is achieved, along with a power-law component with negative normalization to account for deadtime effects \citep[][]{Yadav_2016a, Agrawal_2017}. We find that a model consisting of one power law and five \texttt{Lorentzian} components provides the best description of the PDS in both segments, yielding a $\chi^2$ of 278 for 315 degrees of freedom in Segment 1 and 404.4 for 386 degrees of freedom in Segment 2.

We then compute the cross-spectrum between different energy bands, taking the 4.0–8.0 keV band as the reference. Specifically, we calculate the CS for 8.0–12.0 keV, 12.0–20.0 keV, and 20.0–40.0 keV relative to the reference band. For each case, we fit the real and imaginary parts of the CS simultaneously along with the full-band PDS. To model the real and imaginary parts of the CS, we define additional models in \texttt{xspec}: for each variability component, the real and imaginary parts are described as \texttt{Lorentzian $\times$ cos($\phi$)} and \texttt{Lorentzian $\times$ sin($\phi$)}, respectively, where $\phi$ represents the phase lag between the corresponding energy bands and is treated as a free parameter \citep[see][for more details]{Mendez_2024}. The centroid frequency and width of each \texttt{Lorentzian} are fixed to the values obtained from the PDS-only fit in the full energy band and are tied across the PDS and the real and imaginary parts of the CS during the simultaneous fit.

Figures \ref{Figure: Constant Phase Lag 1 Segment 1} and \ref{Figure: Constant Phase Lag 1 Segment 2} show representative fits for Segments 1 and 2, respectively, where the PDS and the real and imaginary parts of the CS (for 8.0–12.0 keV relative to 4.0–8.0 keV) are fitted simultaneously. Using this procedure, we get the phase lags for all energy bands in both segments. Figure \ref{Figure: Phase Lag Spectrum} shows the observed and model phase-lag spectra for Segments 1 and 2. The observed phase lags are calculated from the real and imaginary parts of the cross-spectrum between the 8.0-12.0 keV energy band and the 4.0-8.0 keV reference band as
\[
\phi(\nu)=\tan^{-1}\left(\frac{\mathrm{Im}[CS(\nu)]}{\mathrm{Re}[CS(\nu)]}\right),
\]
while the model phase lags are derived from the best-fitting models of the real and imaginary parts of the cross-spectrum without directly fitting the phase lags. The derived model reproduces the overall frequency dependence of the observed phase-lag spectra, including the behaviour around the QPO and QPO shoulder frequencies, demonstrating that the simultaneous modelling of the power density spectrum and the real and imaginary parts of the cross-spectrum provides a self-consistent description of the phase-lag behaviour.

\begin{figure*}
    \centering
    \includegraphics[width=0.48\textwidth, height=5.5cm]{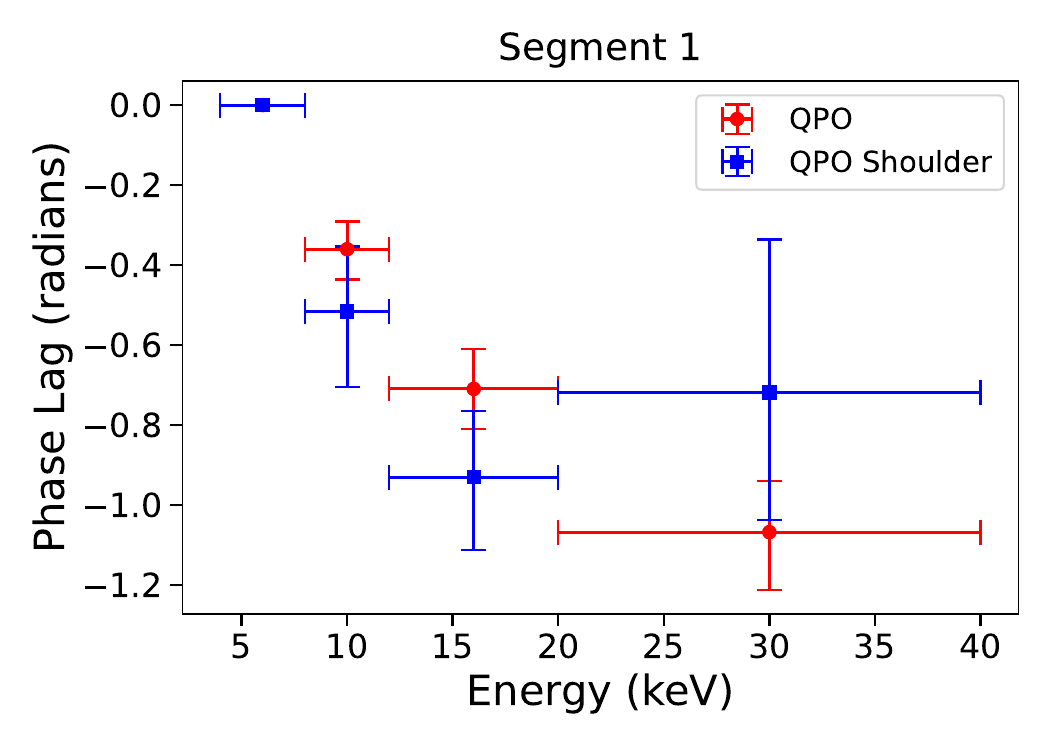}
    \includegraphics[width=0.48\textwidth, height=5.5cm]{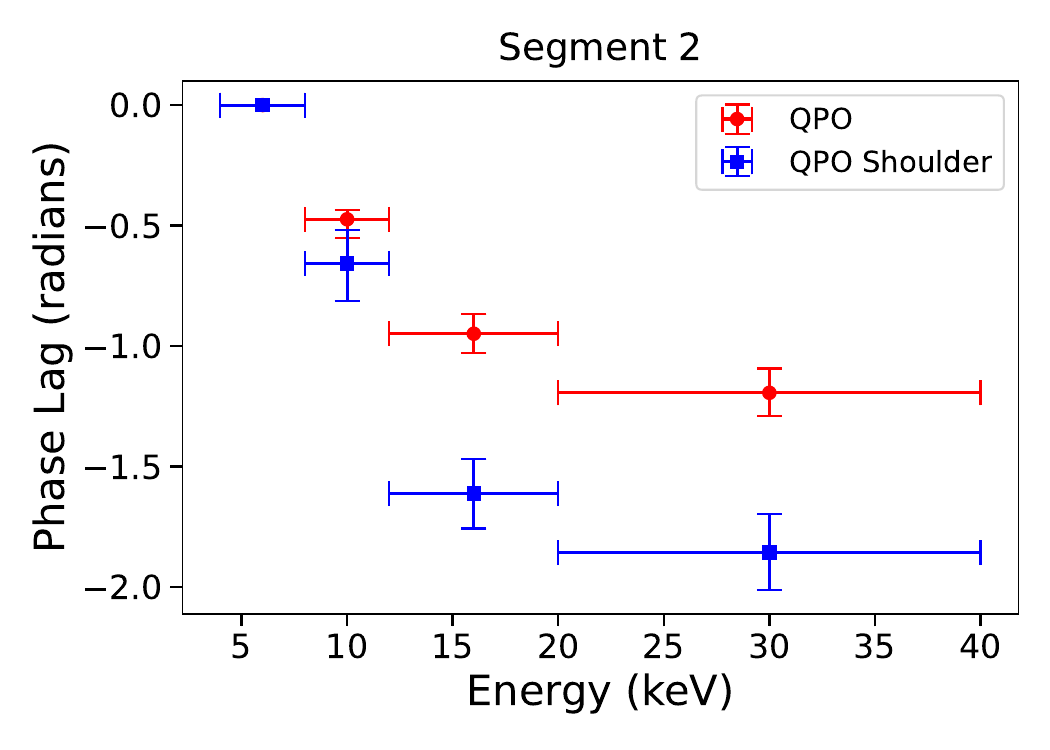}  
    \caption{Phase lag of QPO and QPO Shoulder as a function of energy for Segments 1 (left) and 2 (right) of GRS 1915+105}. In Segment 1, the red and blue points correspond to the Lorentzian components with centroid frequencies of 5.54 Hz and 6.51 Hz, respectively, while in Segment 2 they correspond to the Lorentzian components with centroid frequencies of 6.66 Hz and 7.93 Hz, respectively.
    
    \label{Figure: Phase_Lag_Segments_1_2}
\end{figure*}

Figure \ref{Figure: Phase_Lag_Segments_1_2} shows that both variability components exhibit negative phase lags in Segment 1 (left panel), whose magnitudes generally increase with energy. However, owing to the relatively large uncertainties, the phase-lag spectra are statistically consistent with each other. For example, in the 20-40 keV band, the QPO and QPO shoulder exhibit phase lags of $-1.07 \pm 0.14$ rad and $-0.72 \pm 0.35$ rad, respectively, differing by only $0.35 \pm 0.38$ rad ($\sim0.9\sigma$).

In contrast, Segment 2 (Figure \ref{Figure: Phase_Lag_Segments_1_2}, right panel) reveals a clear separation between the phase-lag spectra of the two components. While both exhibit increasingly negative phase lags with energy, the QPO saturates at $-1.19 \pm 0.10$ rad, whereas the QPO shoulder reaches $-1.86 \pm 0.16$ rad in the 20-40 keV band. The difference of $0.66 \pm 0.18$ rad corresponds to a significance of $\sim3.6\sigma$, providing strong evidence that the two variability components possess intrinsically different phase-lag spectra.

\section{Discussion} \label{Section: Discussion}
We present a detailed energy-resolved analysis of the LFQPO in the BHXB GRS 1915+105 using AstroSat/LAXPC observations. We find that the QPO centroid frequency exhibits a significant increase with energy only when the QPO frequency is $\sim6$ Hz, whereas no measurable energy dependence is detected at $\sim4.5$ Hz (Segment 3), consistent with previous reports \citep[][]{Yadav_2016}. More importantly, we show that the apparent energy dependence of the QPO frequency can be naturally explained by the presence of multiple variability components. While independent fits to the energy-resolved PDSs suggest a single QPO whose centroid shifts with energy, simultaneous fitting demonstrates that the QPO feature is better described by two Lorentzian components compared to a model describing the QPO feature by a single Lorentzian whose centroid frequency and FWHM are energy-dependent (Figure~\ref{Figure: PDS with 2 Lore Segment 2}).

\subsection{Energy-Dependent of QPO Frequency}
We investigate the energy dependence of the QPO feature in GRS 1915+105 and the centroid frequency increases with energy. Similar energy-dependent behaviour has been reported in several BHXBs, where the slope of the frequency-energy relation evolves systematically over a broad range of QPO frequencies \citep[e.g.,][]{Qu_2010, Li_2013, Li_2013b, Yan_2018, Zhu_2024}. In contrast, our observations sample a relatively narrow frequency range ($\sim4.5$-6 Hz), and therefore a much smaller change in the slope is expected. Instead, the energy dependence appears only near $\sim6$ Hz and is absent at lower frequencies, consistent with \citet{Yadav_2016}. Furthermore, the dynamic PDS analysis (Figure~\ref{Figure: DPDS_Segments_1_2}) does not reveal any significant temporal evolution of the QPO centroid frequency, suggesting that the observed energy dependence of the centroid frequency is intrinsic rather than a consequence of time-dependent centroid frequency shift.

Various models have been proposed to explain the origin of LFQPOs in black hole XRBs, including global disk oscillation \citep{Titarchuk_2000}, radial and orbital oscillation \citep{Nowak_1993, Nowak_1994}, accretion flow instability \citep{Nowak_1994}, and drift-blob models \citep{Hua_1997, Bottcher_1998, Bottcher_1999}. More recently, time-dependent Comptonization models have successfully reproduced the observed energy-dependent $frms$ amplitude and phase-lag spectra \citep[e.g.,][]{Karpouzas_2020, Bellavita_2022}, but they do not yet provide a complete explanation for all observed behaviours, particularly the energy dependence of the QPO centroid frequency \citep[e.g.,][]{Qu_2010}. \cite{Eijnden_2016} proposed that differential Lense–Thirring precession of the inner accretion flow could naturally give rise to an energy-dependent QPO frequency, with different disc radii contributing at different energies. To test this toy model, \citet{Eijnden_2016} applied it to RXTE observations previously analysed by \citet{Qu_2010}, \citet{Pahari_2013}, and \citet{Yan_2013}. They showed that the model successfully reproduces the observed energy dependence of the QPO centroid frequency and phase lag, the correlation between the full-band QPO frequency and phase lag, and the decoherence of the QPO over timescales of $\sim5$--10 QPO cycles.

\subsection{Alternative Explanation for Energy-Dependent QPO Frequency}
The energy dependence of the QPO centroid frequency can be naturally explained by the presence of multiple variability components rather than an intrinsic shift of a single QPO frequency. While independent fits to the energy-resolved PDSs describe the QPO feature as a single Lorentzian whose centroid frequency varies with energy, simultaneous fitting of all energy bands (Figure~\ref{Figure: PDS with 2 Lore Segment 2}) shows that a model comprising two Lorentzian components with frequency and FWHM independent of energy provides a better fit. Following \citet{Mendez_2024}, we refer to these components as the QPO and the QPO shoulder, corresponding to the Lorentzian with the lower and higher centroid frequencies, respectively.

The energy-dependent $frms$ amplitude of the QPO and its shoulder further support this interpretation. The two components exhibit significantly different $frms$ amplitude evolution (Figure \ref{Figure: RMS_Segments_1_2}): $frms$ amplitude of the QPO increases with energy and saturates at a relatively lower amplitude, whereas the QPO shoulder shows a steeper rise and reaches a comparatively higher saturation amplitude. A similar result was reported by \cite{Mendez_2024} in their analysis of one of the same RXTE observations studied by \cite{Qu_2010}. In that work, the $frms$ amplitude of the QPO increased from $\sim2.5\%$ and saturated at around $\sim4\%$, while the $frms$ amplitude of the QPO shoulder started from a lower value of $\sim1\%$, rose more rapidly, and saturated at a higher level of $\sim5.5\%$ \citep[see middle panel of Figure 3 in][]{Mendez_2024}.

Adopting the \texttt{constant phase model} \citep[see][for details]{Mendez_2024}, we find that the two components are characterized by negative phase lags but with different behaviours (Figure \ref{Figure: Phase_Lag_Segments_1_2}). In Segment 2, where the constraints on phase lags are robust, both the QPO and its shoulder exhibit negative lags that tend to saturate at different levels. \cite{Mendez_2024} also reported a similar pattern; they showed different saturation levels in the time lag evolution of the QPO and the QPO shoulder \citep[see right panel of Figure 3 in][]{Mendez_2024}.

The simultaneous fit of the energy-resolved PDSs (Figure \ref{Figure: PDS with 2 Lore Segment 2}), together with the corresponding $frms$ amplitude (Figure \ref{Figure: RMS_Segments_1_2}) and phase-lag spectra (Figure \ref{Figure: Phase_Lag_Segments_1_2}), supports that the observed QPO feature is composed of two distinct components, namely the QPO and the QPO shoulder. These two components exhibit different $frms$ amplitude and phase lag behaviours, indicating that they are likely associated with separate variability processes. The large effective area and broad energy coverage of AstroSat/LAXPC, particularly its sensitivity at higher energies, enable the resolution of closely spaced components that might otherwise appear as a single broadened feature. Consequently, the apparent shift in the centroid frequency of the QPO feature with energy arises from changes in the relative contributions of these two components in different energy bands, effectively producing an {\color{red}} energy-dependent evolution of the frequency of the overall QPO feature.

The phenomenological properties of the two Lorentzian components provide additional insight into their possible physical nature. In our analysis, the QPO has a quality factor and $frms$ amplitude of $\sim13.1$ and $\sim3.5\%$ in Segment 1 and those of $\sim13.6$ and $\sim2.25\%$ in Segment 2, consistent with a Type-C QPO \citep[for the classification details of LFQPO, refer to][]{Casella_2005}. The QPO shoulder in both segments, on the other hand, exhibits a considerably lower quality factor ($Q\sim4.2$ and 4.5 in Segments 1 and 2, respectively), although their $frms$ amplitude ($\sim1.6$ and $\sim$3.2\% in Segments 1 and 2, respectively) and frequencies lie within the range commonly associated with Type-B QPOs. Combined with the distinct $frms$ amplitude and phase-lag spectra of the two components, these results suggest that the shoulder represents a separate variability component rather than simply an asymmetric extension of the QPO. Interestingly, \cite{Jin_2026} suggested that a similar shoulder-like feature in Swift J1727.8$-$1613 may be associated with the emergence of a Type-B QPO coexisting with a Type-C QPO. Although the phenomenology observed here is qualitatively similar, particularly the presence of two closely spaced variability components with different energy-dependent temporal properties, the relatively low quality factor of the shoulder and the lack of additional observational signatures characteristic of Type-B QPOs prevent us from drawing such an identification. We therefore interpret the QPO shoulder as a distinct variability component whose physical origin remains uncertain, while noting that the possible coexistence of multiple QPOs is an intriguing possibility deserving further investigation.

\begin{acknowledgments}
We are grateful for the data obtained from the LAXPC instrument on board the AstroSat satellite, which has been used in this study. The analysis has been carried out using the LAXPC software and HEASoft tools. VS, SC, and VJ acknowledge the support and warm hospitality extended by the Inter-University Centre for Astronomy and Astrophysics (IUCAA), Pune, during their visit, which greatly facilitated the completion of this work. VJ acknowledges the support provided by the Department of Science and Technology (DST) under the ‘Fund for Improvement of S $\&$ T Infrastructure (FIST)’ program (SR/FST/PS-I/2022/208). We acknowledge that this work was initiated during the AstroSat $\&$ XPoSat Data Analysis Workshop, supported by the Indian Space Research Organisation, held from June 10 to 14, 2025, and jointly organized by the AstroSat Science Support Cell at IUCAA and Providence Women's College, Kozhikode, Kerala. We also thank Arbind Pradhan (Tezpur University, Assam, India) for his assistance in fitting the real and imaginary parts of the cross-spectrum. VS would like to thank Akash Garg (Inter-University Centre for Astronomy and Astrophysics, Pune, India) and Divya Rawat (Observatoire astronomique de Strasbourg, France) for their insightful and fruitful discussions. We also thank the anonymous referee for constructive comments that have improved the manuscript.
\end{acknowledgments}

\begin{contribution}

All authors contributed equally.


\end{contribution}

%
\facilities{AstroSat (LAXPC)}

\software{LAXPCSoftware, Heasoft, XSPEC}


\bibliography{References}{}
\bibliographystyle{aasjournalv7}



\end{document}